# Applying endogenous learning models in energy system optimization

**Jabir Ali Ouassou[1,*], Julian Straus[1], Marte Fodstad[1], Gunhild Reigstad[1], Ove Wolfgang[1]**

[1] SINTEF Energy Research
* Correspondence: jabir.ouassou@sintef.no

**Abstract:**
Conventional energy production based on fossil fuels causes emissions which contribute to global warming. Accurate energy system models are required for a cost-optimal transition to a zero-emission energy system, an endeavor that requires an accurate modeling of cost reductions due to technological learning effects. In this review, we summarize common methodologies for modeling technological learning and associated cost reductions. The focus is on learning effects in hydrogen production technologies due to their importance in a low-carbon energy system, as well as the application of endogenous learning in energy system models. Finally, we present an overview of the learning rates of relevant low-carbon technologies required to model future energy systems.

**Keywords:**
Learning by doing; Learning curve; Learning rate; Endogenous learning; Energy system models.

## 1. Introduction

It is by now well-known that conventional energy production based on fossil fuels cause problematic CO2 emissions which contribute to global warming. Avoiding a climate crisis requires a worldwide effort to reduce such greenhouse gas emissions, and the EU has set its target to be net-zero emissions by 2050 [1]. Since energy production and consumption (including transport) accounts for approximately 80 % of the greenhouse gas emissions in the EU [2], most of this burden falls to the energy sector. Some reduction is expected from energy efficiency increases and the circular economy, but the majority must likely come from investments in novel energy technologies: renewable electricity, biofuels, hydrogen as an energy carrier, carbon capture and storage, *etc*. In this context, energy system models, which attempt to forecast and optimize the entire energy system in, *e.g.*, the EU, are powerful tools for guiding policymakers and minimizing transition costs. Moreover, their cost predictions can be useful for analyzing the competing technologies within a sector. As examples, the TIMES model generator [3] is used globally in providing guidance regarding future policies while the PRIMES model [4] is used by the European Commission in long-term strategies like "A Clean Planet for all" [5].

One fundamental challenge for such energy system models is that one requires estimates for the costs of every relevant energy technology within the investigated timeframe. However, the investment required to, *e.g.*, build and operate one new power plant can change drastically from 2020 to 2050: costs typically decrease as a new technology matures and becomes more widely deployed, but can also increase due to changes in raw materials or regulations resulting in more stringent safety protocols. These costs are especially important for renewable energy technologies as their cost of electricity is dominated by the capital costs and are not affected by fossil fuel prices. Samadi [6] provides a broad overview of the factors that typically influence the costs of electricity generation. He groups these into four main clusters: learning and technological improvements, economies of scale, changes in input factor prices, and social and geographical factors. Within the context of energy system models, learning and technological improvements and economies of scale are especially important as these can be directly affected by variables in the model.

Changes in technology costs due to learning effects can be implemented either exogenously [7] or endogenously [8] in an energy system model [9]. Exogenous learning means that one models the technology cost purely as a function of time, independent of any investment choices made during the energy system optimization. In other words, the technology cost forecast can be regarded as an input to the model. They are frequently obtained from in-depth analysis of the individual energy technologies. Conversely, endogenous learning means that the costs are assumed to be some function of the prior investments, and changes dynamically as the energy system optimization routine explores different investment choices. Thus, the technology cost forecast is then an output from the energy system optimization. This is clearly a more realistic scenario, and avoids some pitfalls associated with the exogenous learning models. For instance, in the exogenous case an investment algorithm may choose to delay all investments for a decade since the technology cost will have decreased by then, while in the endogenous case the costs of a new technology do not decrease unless someone invests in the technology. The main drawback of this strategy is that endogenous learning causes the optimization problem to become nonlinear, requiring more advanced and computationally expensive solution algorithms than exogenous learning. Furthermore, problems may arise with learning occurring outside the investigated bounds of the energy system models resulting in higher costs within the model than in reality. This necessitates adjustments of the learning rates.

To provide a realistic pathway to a zero-emission future, energy system modeling including learning effects can be a useful planning tool. Moreover, hydrogen is increasingly considered as an important part of such a zero-emission energy system. However, there is currently a lack of reviews that summarize the available research on learning effects in hydrogen production, presenting a difficulty for incorporating these effects into energy system models. The aim of this publication is precisely to investigate learning rates for hydrogen production and to focus on the implementation and implications of the endogenous learning approach in energy system models. The review is structured as follows. We provide a conceptual introduction to endogenous learning models in section 2, focusing on learning-by-doing and learning-by-research effects in particular. In section 3, we present the results of a thorough literature review, where we investigated what learning rate data for energy technologies are publicly available. A key focus in this area is to also include technologies outside the typical electricity generation technologies. Finally, we conclude with a discussion of problems with endogenous cost reductions in Section 4.

## 2. Cost development in technologies

There exist different strategies for estimating the future costs of an energy production technology [10]. These can broadly be categorized as "bottom-up estimates", based on state-of-the-art research and engineering combined with detailed domain knowledge for the calculation of an *n*'th-of-a-kind (NOAK) plant; "top-down estimates", based on extrapolating purely empirical trends; as well as combinations of these, as suggested by Rubin [10] and discussed by Roussanaly *et al*. [11]. The following sections provide an overview of the mathematical concept of top-down estimates via learning curves.

*2.1. Introduction to learning curve models*

The concept of a learning curve or experience curve was originally introduced by Wright [12]. He observed that every time the total number of aircraft that had been produced doubled, the number of person-hours needed to produce one more aircraft had decreased by 20 %. This effect is now known as learning-by-doing in economics. The same trend was later found to hold not just within one company, but for entire industries; and not just in manufacturing, but also for other industries such as energy production [13].

The method originally introduced by Wright is now known as the one-factor learning curve model. Mathematically, it is formulated as follows [14]:

$$C(x) = C_0 \left(\frac{x}{x_0}\right)^{b_{\text{lbd}}}. \tag{1}$$

In Wright's example, $C(x)$ measured the cost of producing one airplane after a total of $x$ airplanes have been produced, while $C_0$ and $x_0$ were the corresponding values at some earlier reference time $t_0$. The exponent $b_{\text{lbd}}$ is usually parametrized via a *learning rate* $\text{LR} = 1 - 2^{b_{\text{lbd}}}$, which describes the cost reduction obtainable by doubling $x$.[1] In Wright's example, LR = 20 %; generally, it is determined by fitting historical data to Eq. (1). See Figure 1: Conceptual illustration of a standard one-factor learning curve. As for Wright's airplanes, we assume a learning rate LR = 20 %. The blue curve shows how the unit cost decreases as a function of installed capacity, while the green lines emphasize that the cost $C$ is reduced by 20 % after successive doublings of capacity $x$. for a visualization of this one-factor learning curve. Note that this learning curve is only valid for technologies that have already been commercialized; until that point, the cost does not generally follow this curve, and may initially even increase with time [10,11]. To remedy this, some sources use a constant estimate $C(x) = C_0$ when modeling capacities $x < x_0$ [15], where $x_0$ is an estimated production threshold before learning starts.

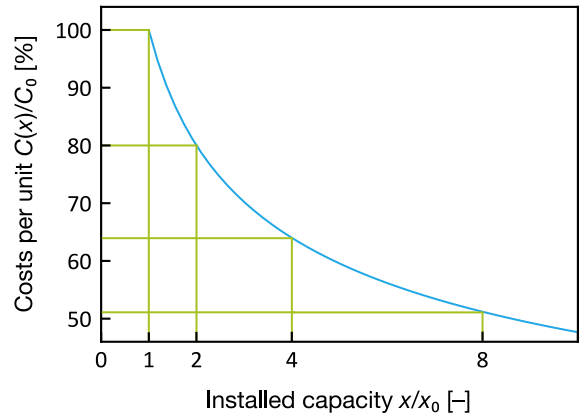

**Figure 1:** Conceptual illustration of a standard one-factor learning curve. As for Wright's airplanes, we assume a learning rate LR = 20 %. The blue curve shows how the unit cost decreases as a function of installed capacity, while the green lines emphasize that the cost $C$ is reduced by 20 % after successive doublings of capacity $x$.

The simple one-factor approach outlined above may result in the incorporation of factors independent of the installed capacities into the learning rates. This may result in both over- and underestimation of the associated learning rates. For a concrete example: both private companies and public institutions spend a lot of resources on research and development (R&D), which should reduce technology costs over time. This is often called *learning-by-research* and is a distinct effect from the learning-by-doing discussed above [9], as it scales with the research budget and not the production or capacity. However, if one fits a one-factor model to the historical cost development, then all cost reductions due to R&D investments will be attributed to learning-by-doing in the model, resulting in an overestimation of the learning-by-doing effect. If the research budget and production rate are strongly correlated, this is mainly a philosophical issue; in that case, the one-factor model would retain its predictive power as one extrapolates from empirical data. However, if the research budget and production rate change differently with time, one might expect more accurate predictions if the learning-by-doing and learning-by-research effects are modeled separately. Such an extended *two-factor model* can be expressed as [9]:

$$C(x,y) = C_0 \left(\frac{x}{x_0}\right)^{b_{\text{lbd}}} \left(\frac{y}{y_0}\right)^{b_{\text{lbr}}} \qquad (2)$$

---

[1] If we have an initial capacity $x_0$, and double this $n$ times to get $x = 2^n x_0$, then a one-factor learning curve yields $C = C_0 (x/x_0)^{b_{\text{lbd}}} = C_0 (2^n)^{b_{\text{lbd}}} = C_0 (2^{b_{\text{lbd}}})^n$. Substituting in the definition $2^{b_{\text{lbd}}} = 1 - \text{LR}$, we get $C = C_0 (1 - \text{LR})^n$. This shows that the cost indeed decreases by a fraction LR for each of the $n$ doublings in capacity that occurs.

where $y$ corresponds to the R&D spending and $b_{lbr}$ to the corresponding learning-by-research parameter. Rubin *et al.* [9] concluded based on a literature review that R&D can significantly reduce the costs in all stages of technology development. Following the same pattern as above, one can continue to extend the learning curve concept to an arbitrary *multi-factor model* with independent learning rates for each factor included. While multi-factor models are theoretically appealing, in practice one-factor learning curves are more used due to a lack of available data [9].

Both the one-factor and two-factor learning curves can be expanded using *component-based learning curves*, also known as *composite learning curves*. In these models, a "unit" is decomposed into its constituent parts, and each of these parts is then allowed to evolve according to a different learning rate. As an example, consider hydrogen production using natural gas reforming with carbon capture and storage. The overall system consists of a reforming section, a hydrogen purification section, a $CO_2$ capture and processing section, a $CO_2$ transport section, and a $CO_2$ storage section. Each section has its own maturity and potential cost reductions through increase in production capacity. This can either be solved using a *composite learning rate*, where the learning associated with each component is aggregated into a single effective learning rate, or by using individual learning rates. The former may result in wrong assumptions on the overall learning rate through an omitted variable bias [16]. The latter requires the utilization of composite learning curves, which for a one-factor model is given by [9,17]:

$$C(x_1, \ldots, x_N) = \sum_n C_n(x_n) = \sum_n C_{0,n} \left(\frac{x_n}{x_{0,n}}\right)^{b_{lbd,n}} \quad (3)$$

where $C$ is the total cost, $C_n$ is the cost of the $n$'th component, and $x_n$ is a corresponding production or capacity. This approach also makes it possible to model spillovers from other technologies, by referring to the same component $x_n$ in multiple cost calculations. One example of such spillovers is given by fuel cell electric vehicles and electric vehicles where battery and electric drive train costs are common in both type of vehicles [18]. Another relevant example is photovoltaics, where much of the historical improvements can be attributed to technology spillover from the semiconductor electronics industry.

The simplest version of the component approach discussed above is to only distinguish between components with significant learning effects ($b_{lbd,i} \gg 0$) and negligible learning effects ($b_{lbd,i} \approx 0$). The former is then used to calculate a composite learning rate $b_{lbd}$ for the fraction $\alpha$ of the initial costs that have significant learning effects. The remaining fraction $1 - \alpha$ of the initial costs will remain constant, and thus over time become the dominant contribution to the total cost. Mathematically, such a model can be written as:

$$C(x) = C_0 \left[(1 - \alpha) + \alpha \left(\frac{x}{x_0}\right)^{b_{lbd}}\right] \quad (4)$$

In general, a combination of multi-component and multi-factor analyses yields a sum over contributions from different components, where the cost of each component is now a product of cost-decreasing factors [19]. Each of these factors (learning-by-doing, learning-by-research, *etc.*) has a separate learning rate that must be estimated based on data. Estimating the uncertainty can also become more complex since the learning rates are often correlated, not independent. A key problem of both composite learning curves and multi-factor approaches is thus the availability of data that can be used for obtaining these learning parameters without overfitting, as well as the quality of the available data.

A recent review by Thomassen *et al.* [20], which focused on using learning curves for technology assessment and how to incorporate environmental concerns in these models, also touched upon multi-component multi-factor learning curves. They formulated a set of recommendations for how to best apply such learning curves in practice. These included the following suggestions: combining learning rates at the component and product levels; combining extrapolated empirical data with expert estimates; and using a tier-based method with quality criteria to evaluate the learning curves.

In practice, there are many ways to define $C$ and $x$ in Eq. (1). In the context of energy system modeling, the *cost variable C* usually refers to a relative cost (*e.g.*, €/kW or €/unit) or levelized cost (LCOE), while the *experience variable x* refers to installed capacity (GW), installed number (units), or total production (TWh of energy production). Learning curves are also used to describe other developments, such as operating efficiency [21] and consumption of individual input factors in production [22]; however, this study is limited to the cost-related use of learning curves due to its prevalence in energy system models.

## 3. Literature review related to learning-by-doing

Learning-by-doing effects have been extensively studied for the technologies present in an energy system. It is not surprising that especially onshore wind and solar photovoltaics have received significant attention due to their importance in a low-carbon energy system and their high growth rate in recent years. However, as illustrated in Figure 2, there is a significant spread between the learning rates reported in the literature, although statistical analysis narrows down the intervals. This holds even if there are a large number of publications analyzing a specific technology. Even so, the uncertainty in the learning rate can often be as large as the value of the learning rate, which can have drastic effects on outcomes of energy system models given that the learning rate appears in the exponent of the learning curve expressions. This underlines the importance of an explicit treatment of learning curve uncertainty, *e.g.*, through the application of sensitivity analyses.

The learning rates uncovered in this literature review—which were used as a basis for Figure 2 and the following subsections—are presented in tabular form in Appendix C. More in-depth discussion of the literature on electricity-related energy technologies and carbon capture and storage is delegated to Appendices A and B. In the rest of this section, we summarize the results of the literature review, focusing on learning rates in (i) hydrogen production technologies and (ii) their application in energy system models.

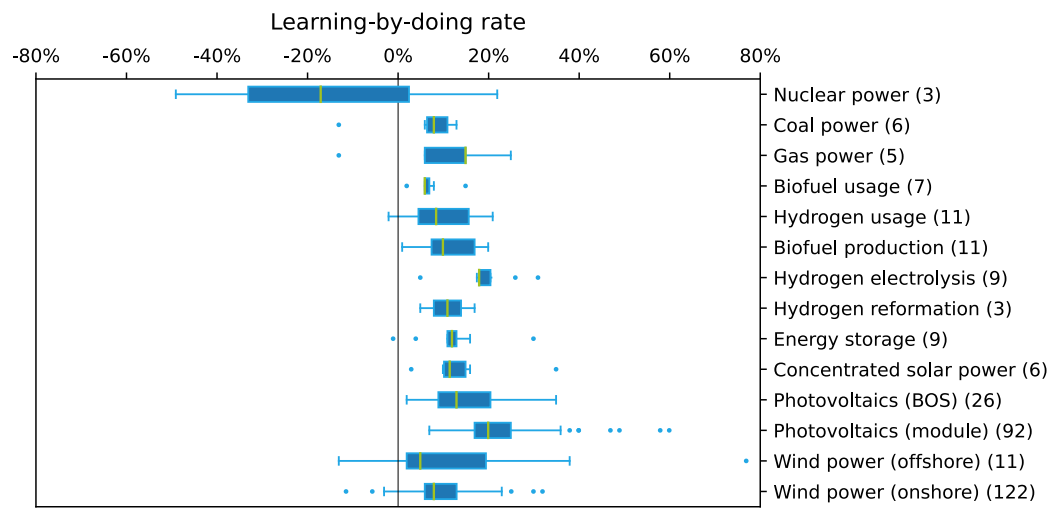

**Figure 2:** Statistical box plot showing the distribution of learning-by-doing rates uncovered by our literature review. The green vertical lines and surrounding blue boxes show the median and inter-quartile range, respectively. As in standard box plots, the whiskers show the distribution of remaining data points that are not considered outliers, while the dots denote the outliers.

*3.1. Learning-by-doing in hydrogen production*

Hydrogen is increasingly considered to be an important energy carrier in a low-carbon energy system [23,24]. Hence, it is important to also include hydrogen production with learning effects in an energy system model to avoid a potential bias. Most studies on learning effects in hydrogen production focus on electrolysis due to its low deployment to date. Hence, significant cost reductions can be expected with increased deployment.

Learning effects for hydrogen production from natural gas has been less investigated. One reason is that much of this hydrogen goes to refineries, ammonia production facilities, and methanol production facilities [25]. In these processes, hydrogen is directly utilized while the composition of the feedstock may vary. These production processes differ significantly, making it complicated to obtain reasonable learning rates. For instance, methanol plants do not maximize their hydrogen yield, as the ideal ratio between hydrogen and carbon monoxide is two for methanol synthesis [25]. The challenges of estimating learning effects for such hydrogen production processes is elaborated in Rubin *et al.* [26]. They note that complicating factors include differences in feedstocks, plant size, desired hydrogen purity, the possibility of steam export, and economic factors.

One notable exception, where learning rates for hydrogen production from steam–methane reforming (SMR) and carbon capture were estimated, is the work by Rubin *et al.* [15,26]. SMR learning rates were estimated based on the unit price for hydrogen assuming a fixed share between energy input, capital charges, and operation and maintenance (O&M) costs, with an annual reduction in natural gas consumption of 1.1 %. The installed capacity here corresponds to the total amount of produced hydrogen worldwide. They concluded with a learning rate of 27 % both for the capital charges and O&M costs.

In another study, Schoots, *et al.* [27] investigated learning effects with respect to hydrogen production technologies. Their study used cost data that goes back to the 1940s and focused on SMR, coal gasification, and alkaline electrolysis. Partial oxidation of oil and naphtha was not included due to limited data quality and the large variety of feedstock. The production capacity was scaled to avoid confusing learning-by-doing and economy of scale. They found significant learning rates of 11±6 % and 18±13 % for the investment costs of SMR and electrolysis, respectively. Important factors preventing even higher learning rates were improved energy efficiency, stricter environmental regulations, and potentially stricter hydrogen purity requirements. All these effects increase capital costs. The production costs were however insensitive to learning effects, as most of the production costs are related to energy input in the form of natural gas, coal, or electricity. Further variable costs like insurance or personnel do not show learning at all. These results cannot directly be compared to the results by Rubin *et al.* as the basis for the learning rate is different and the study accounted for the economy of scale in hydrogen production.

Schmidt, *et al.* [28] performed an expert elicitation study for the future costs of electrolysis and compared the potential future costs proposed by domain experts to the ones obtained via learning curves. They investigated alkaline electrolysis (AEC), proton exchange membrane electrolysis (PEMEC), and solid oxide electrolysis (SOEC). They used a learning rate of 18 % from proton exchange membrane fuel cells for PEMEC due to a lack of data. Similarly, solid oxide fuel cells with a learning rate of 28 % were used as surrogates for SOEC. The reason for using learning rates for fuel cells was their similarity to electrolysers. With two different deployment scenarios, they analyzed the future cost predictions by the experts. These were mostly in line with the learning rates and their uncertainties—except for the AEC, where the experts predicted lower costs.

Böhm *et al.* [19] analyzed electrolysers via composite learning curves—specifically, via a modified version of Eq. (3). Furthermore, spillover effects between electrolyser types were included for power electronics and gas conditioning. Their analysis showed that a component-based approach reduced the learning rate as production increased. This was because the initially most expensive components also had the highest learning rates, so increased production shifted the cost distribution towards components with lower learning potential. For example, bipolar plates with a learning rate of 18 % represented 51 % of the initial cost of a PEM electrolysis cell; but after 1,000 times more cells had been produced, this share had dropped to 24 %. The learning rates reported for AEC, PEMEC, and SOEC were 19.5 %, 17.5 %, and 20.5 % respectively at the first stage of deployment. Except for SOEC, these values are similar to the ones reported by Schmidt *et al.* [28]. Krishnan, *et al.* [29] reported a learning rate of 16±6 % for AEC, which is also similar to the other values.

Haltiwanger, *et al.* [30] investigated a different production route through a Zn/ZnO thermochemical cycle and concentrated solar power (CSP). As there exists no reliable cost data for the reactor for the cycle, SMR data from Rubin *et al.* [15] was used as a surrogate. The data for the CSP section was obtained by splitting the cost for a CSP plant into a steam cycle section and the solar components. The cumulated learning rate was calculated as 13–23 % with the levelized cost of hydrogen as the cost basis and the cumulative hydrogen production as the capacity basis. This study was later extended to compare the solar thermochemical cycle to photovoltaic with electrolysis with composite learning rates allowing different growth rates in photovoltaics and electrolysers, as well as the Zn/ZnO cycle and the CSP plant [17]. In this study, the thermochemical cycle was modelled with a learning rate of 19±8 %. This was based on a prior study of 108 different data sets from 22 industrial sectors [31], which concluded that their learning rates were normally distributed with a mean of 19 % and standard deviation of 8 %.

*3.2. Application of learning-by-doing in energy system models*

As shown in Section 2, the learning-by-doing requires the implementation of nonlinear, nonconvex functions. Most energy system models are either formulated as linear problems or mixed-integer linear problems. The former formulation does not allow learning curves, and must rely on exogenous learning approaches. Mixed-integer linear problems allow implantation through an approximation as piecewise linear functions for the cumulative cost as is the case, *e.g.*, in the Endogenous Technological Learning (ETL) extension for the TIMES model [3]. Using the cumulative investment costs furthermore avoids problems with bilinear terms in the objective function. However, this translates the original linear problem into a mixed-integer linear problem, which drastically increases the computational complexity. Therefore, in general, learning-by-doing is only implemented in models focusing on specific sections of the energy system, *e.g.* the power supply section without geographical scope or simple technologies with geographical scope, and not in models focusing on the overall energy system like TIMES. For an overview of the individual models, we refer to Junginger, *et al.* [32] and Heuberger, *et al.* [14]. From here on, we focus on the impact of endogenous learning on the energy system model results.

Heuberger *et al.* [14] developed a power capacity expansion model with endogenous learning called ESO-XEL. Technology learning was implemented in the MILP framework as piecewise linear functions *via* the cumulative cost and not the unit cost, as described in detail by Gómez [33]. The utilization of cumulative costs avoids non-linearities in the cost function. The aim of the paper is to analyze capacity expansion in the United Kingdom for a reduction of $CO_2$ emissions by 2050. The simultaneous global expansion of generation technologies was identified as a problem with a local capacity expansion model using endogenous learning. Hence, they modified the learning rate to account for the forecasted capacity expansion on a global level. This is especially important for offshore wind expansion due to high initial costs preventing its widespread implementation. However, the installed capacity in general also depends on the learning rate. Specifically, assuming learning for carbon capture increased the deployment of fossil power plants with carbon capture compared to the approach with constant prices. This was mostly due to the low installed numbers to date, counterweighing the smaller learning rate. Heuberger *et al.* [14] also included a constraint on the maximum capacity expansion for a technology per year, *i.e.* an implementation speed constraint. This approach avoids an issue where a model may choose an unrealistically large investment in a single technology to minimize costs.

Daggash and Mac Dowell [34] utilized the ESO-XEL model for analyzing how net-negative emissions can be achieved in the power system and what the most cost-effective approach was. To this end, $CO_2$ direct air capture (DAC) was included with an assumed learning rate of 9 %. Despite having a moderate learning rate and low existing capacity, which should imply a large potential for learning by doing, DAC was only utilized in net-negative scenarios when biomass + CCS was insufficient to remove $CO_2$ from the atmosphere. Here, the high learning effects did not outweigh the high operational and investment costs of direct air capture. Chen, *et al.* [35] used a similar approach to analyze the impact of bioenergy with CCS in the United Kingdom. As they had a geographically discretized model, more knowledge related to transport chains could be incorporated. Their subsequent sensitivity analysis showed that the learning rate had a significant impact on the costs of removing $CO_2$ from the atmosphere. However, due to the focus on biomass supply chains and inclusion of only two technologies, it was not possible to distinguish the results from simulations without learning except for the costs.

Handayani, *et al.* [36] implemented endogenous learning in the LEAP model for analyzing the capacity expansion for the Java-Bali system in Indonesia. Contrary to Heuberger *et al.* [14]., they found that learning was reduced over the course of the model in fixed time frames. In the renewable energy scenario, the maximum amount of fossil energy used was limited. This implied a constraint on how much low-carbon technologies must be included as a minimum. Utilizing learning-by-doing influenced the distribution of technologies, but not the total amount of low-carbon energy sources, as coal without carbon capture was still cheaper than renewable energies. Hence, their model selected the the minimum allowed number of low-carbon energy sources.

Cerniauskas, *et al.* [37] investigated a supply chain for hydrogen from electrolysis with utilization in both transport and industry in Germany. Learning rates were assumed for both the electrolysers (20 %, derived from several literature sources, among others Refs. [27,28]), and the hydrogen refueling station (learning rate of 6 %, based on data from AirLiquide). However, the learning rates only affected the overall costs as there was no comparison with different technologies. Hence, conclusions about the effect of learning rates with respect to which technology is favorable could not be drawn.

The National Energy Modelling System (NEMS) [38] uses endogenous learning with different learning stages [39]. They differentiated between a revolutionary stage (up to 3 doublings in capacity), evolutionary stage (3–8 doublings in capacity), and a conventional stage (after 8 doublings). Each subsequent stage had a reduction in learning rate due to the maturity of the technology. Unfortunately, it is not possible to investigate the impact of the implementation on the results of the system, as there is no study analyzing the development of the energy system with and without endogenous cost reductions. Similarly, the Regional Model of Investments and Development (REMIND) [40] used endogenous learning curves in a nonlinear model, although only for solar PV, CSP, wind power, a generalized storage unit, as well as hybrid, electric and fuel cell vehicles. When comparing different integrated assessment models, it was highlighted that the usage of learning curves resulted in different primary energy usage [41]. However, due to the limitation to certain technologies, it may also have included a bias for the technologies with learning.

The Horizon 2020 project REFLEX [42] implemented learning-by-doing within three different models to analyze the development of the future European energy system. Within the model FORECAST, learning rates for heating applications showed reduced costs, but its influence on the installation rate was limited, highlighting alternative reasons for choosing a specific heating solutions [43]. Contrary, within the model PowerACE, learning rates affected the investment in flexible power generators, resulting in a switch from gas turbines and compressed air storage to batteries if the learning rate for batteries was increased. Similarly, the model ELTRAMOD reduced the installed capacity for gas power with CCS if it was assumed that the CCS components did not experience learning.

## 4. Discussion of the implementation of learning-by-doing in energy system models

Despite their advantages for improving capacity expansion and energy system models, both the concept of learning rates in general and learning-by-doing in particular face criticism. Samadi [16] divides this criticism into three categories: criticism of the theoretical approach, criticism of the empirical data used, and criticism of the use of learning rates. In addition to these three categories, the following section will discuss the form of learning curves and how implementation speed limitations may affect the choice of technologies.

### 4.1. Different forms than standard learning curves

All publications reviewed so far on learning rates and their implementation in energy system models used standard learning curves, *i.e.*, they modeled cost as a power function of the installed capacity. However, Yeh and Rubin [13] argue that these types of learning curves may not accurately model the costs of new technologies. Using as examples flue gas desulfurization and selective catalytic reduction in coal power plants, they show that learning curves behave differently in the early stage. Due to an accelerated deployment of these technologies, the improvements by learning-by-doing were reduced in this stage followed by an increase in the learning rate in subsequent capacity expansion. This would result in an *S*-shaped curve for the learning rate. For these two technologies, we can furthermore observe a price increase in the early stage due to problems in scaling up new technologies and wrong estimates. This is irrelevant for commercialized technologies like wind power and photovoltaics which already have a high existing capacity. However, it may result in difficulties estimating the real cost for, *e.g.,* large-scale hydrogen production technologies, as both natural gas reforming with carbon capture and large-scale electrolyser deployment can be expected to still be in this pre-commercial phase.

A second aspect of *S*-shaped learning curves is that the learning rates level off at a high installed capacity. This seems logical as prices cannot decrease indefinitely, in which case they would eventually approach zero. Narbel and Hansen [44] utilize a reduction in the learning rate based on the installed capacity. This is modelled as

$$\text{LR}(x) = \text{LR}_0 (1-d)^{\log_2(x/x_0)} \qquad (5)$$

where $d$ corresponds to a diminishing rate defined exogenously. This function accounts for reduced learning in future implementations. They use this approach to estimate the costs of different energy scenarios. Correspondingly, the overall costs depend on the chosen diminishing rate. The resulting learning effect may be more reasonable as infinite learning would be unrealistic. On the other hand, the diminishing factor is an additional uncertain model parameter as it is *a priori* hard to know how much learning is ultimately achievable. This approach can be seen as similar to the application in NEMS [39].

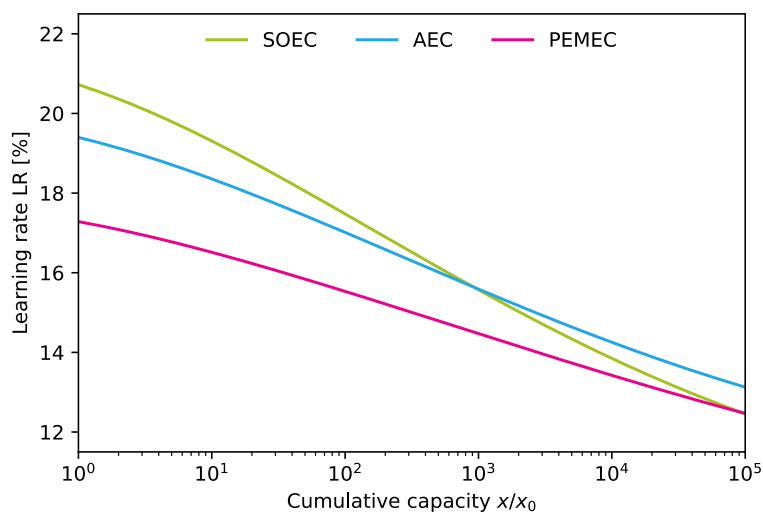

**Figure 3:** Diminishing learning rates for electrolyser stacks for increasing capacity. Reproduced based on Böhm et al. [19].

As outlined in Section 3.1, Böhm *et al.* [19] used a component learning curve approach which resulted in diminishing electrolyser learning rates as the capacity increased. Intuitively, this occurs because the components with a large learning rate must diminish in cost fastest, causing technologies with lower learning rates to become a larger fraction of the total cost. Figure 3 shows this reduction in the overall learning rate for the electrolyser cell stacks. The advantage of the bottom-up modelling from components is a better description of the future costs through accounting for the different learning rates in the individual components. Compared to the approach of Narbel and Hansen [44], no assumption for the reduction in learning is applied. This corresponds to the composite learning curve described in Eq. (3). However, this approach requires detailed knowledge of both the cost distribution for a technology and the corresponding learning rates and still allows a continuous reduction in costs. A similar approach was chosen by Elshurafa, *et al.* [45] for calculating learning rates for solar PV systems as historical data showed a shift in the contribution of the balance of system and the module price on the total price of PV systems.

In their analysis of the Java-Bali electricity system, Handayani *et al.* [36] considered four ten-year time periods and assumed different learning rates for each period. They considered three different scenarios with different assumptions for these learning rates, but did however not compare the study outcome to cases with constant learning rate across all time periods. The chosen scenario influenced the distribution between the different renewable energy technologies. This approach may however result in more errors as the prediction of the reduction in the learning rate is uncertain. Note that the concept itself is similar to the one utilized in NEMS [39]. However, a key difference is that NEMS implements a reduction based on the installed capacity and not the time.

As costs cannot fall indefinitely, another approach is to define a minimum cost which the learning curve is assumed to converge to. The motivation is that material and production costs will at one point be the limit. For instance, Viebahn, *et al.* [46] assume a minimum price for the power block in a CSP plant. A similar approach is taken by Schmidt *et al.* [28] to calculate the minimum prices of electrical energy storage. Similarly to the previously mentioned problems, the exact value of the minimum cost value is difficult to estimate and may lead to wrong interpretations of energy system model results.

*4.2. Implementation speed constraints*

An *implementation speed constraint* refers to a maximum capacity expansion per time unit. A physical analogy is the maximum production capacity of, *e.g.*, batteries. Learning-by-doing favors early cost reductions through investing heavily in technologies with a large learning rate and a low installed capacity [14]. Hence, it may result in unrealistic composition of the energy system if the implementation speed is not restricted. The estimation of the implementation speed for the foreseeable future is possible due to the knowledge of investment in production capacities and the knowledge of current production capacities. The estimation of the implementation speed in, *e.g.*, the 2040s is on the other hand difficult due the potential accumulation of production capacities and a correspondingly high uncertainty. This is a common issue with endogenous learning models: one may end up replacing an exogenous cost with an exogenous capacity expansion constraint, thus substituting one exogenous assumption or prediction for another. This is also pointed out by Lolou *et al.* [3], who mentions that frequently this bound is an active constraint, raising questions around the advantages of endogenous compared to exogenous learning.

Heuberger *et al.* [14] estimated the implementation speed as a building rate from historic data showing how fast technologies can be implemented. The building rates of technologies not deployed on a large scale in the UK were estimated from comparable European countries. Due to the uncertainty associated with the building rates, scenarios were conducted with high and low building rates. These scenarios show the impact of the chosen rates on the final power generation and have a significant impact on the cumulative total system cost. The other studies in Section 0 do no not mention such a constraint. This can be potentially explained by the limited number of technologies utilized, and hence, avoiding overinvestment in a single technology for satisfying demand constraints.

## 5. Conclusion

We have reviewed the available data on learning rates for technologies that are expected to be important in future energy systems. Compared to existing reviews, our main focus has been on hydrogen production technologies, how endogenous learning models are applied in energy system models, and how the utilization of endogenous learning may affect the model results. Current implementation of endogenous learning models largely focuses on models representing the power sector for individual countries. Certain publications narrow the focus further down to individual value chains. Here, the key aim of endogenous learning is to improve future cost estimates.

However, the application of endogenous learning is still limited in energy system models. Some notable challenges to incorporating endogenous learning in such models include the uncertainty in the learning rate values, how to include global cost reductions within a local model and obtaining reasonable estimates for the implementation speed. Considering all these factors, the advantage of using the more realistic approach of endogenous cost reduction may not always outweigh the associated computational costs.

**Author Contributions:** Conceptualization, O.W.; methodology, O.W.; investigation, J.S., J.A.O., M.F.; data curation, J.S., J.A.O., M.F.; writing—original draft preparation, J.S., J.A.O.; writing—review and editing, J.S., J.A.O., M.F., G.R., O.W.; visualization, J.S., J.A.O.; supervision, O.W.; project administration, G.R.; funding acquisition, G.R. All authors have read and agreed to the published version of the manuscript.

**Funding:** The authors acknowledge the *Hydrogen for Europe* study for funding the scientific work which the current publication is based on.

**Acknowledgments:** The authors acknowledge Laila Aksetøy for assisting with finding the initial references used for the literature review.

**Conflicts of Interest:** The authors declare no conflict of interest.

**Appendix A: Learning by doing in electricity generation and storage**

Rubin, Azevedo, Jaramillo and Yeh [9] provide a comprehensive review on the application of learning rates in the electricity generation sector with historical data. Learning rates are presented for in total 11 electric power generation technologies, including fossil plants. The data is regional and also includes learning-by-research in some models. They report a large spread in the different estimated learning rates in renewable energy technologies, while fossil fuel plants have both a reduced learning rate and a reduced spread.

Similarly, Samadi [16] reviewed 67 studies related to the learning rate. A key outcome of the review is that small-scale generation technologies (*e.g.*, wind and PV) tend to have higher learning rates than large-scale technologies (*e.g.*, coal or nuclear). He argues that the reason is that small scale units can be mass manufactured in similar forms in a central production facility resulting in standardization and improvement in the production process, similar to the discovery of learning effects in plane manufacturing by Wright [12]. In contrast, large-scale plants require that the majority of the construction is conducted on site, reducing the impact of improved manufacturing processes. Similarly, large-scale technologies are built in smaller number of units, reducing the impact of improvement in manufacturing processes. In the rest of this section, we proceed to discuss publications not mentioned in the reviews by Rubin and Samadi in some more detail.

Due to the large deployment of solar energy in recent years, learning effects play a significant role in their costs. The learning rate of PV modules has been thoroughly studied, while learning rates for the other components necessary for a solar power system, jointly denoted balance-of-system (BOS), are increasingly receiving attention [45,47]. Some authors also identify the learning rate of the inverter separately from BOS [48]. While PV-module learning rates are frequently assumed to be global [9], D'Errico [47] suggests a combination of national and global drivers for BOS and Duke, *et al.* [49] emphasize local drivers for BOS learning rates. Rubin, Azevedo, Jaramillo and Yeh [9] report a mean learning rate of 23 % and a four-fold range when reviewing learning rates from one-factor LBD models for the module costs. Later one-factor LBD studies [21,50-53] show similar results, ranging from 8.10 % [50] to 39.8 % [21]. Ding, *et al.* [54] develop a one-factor LBR model and find global learning rates of 48.5 % and 35.9 % without and with a two-year time lag, respectively. One-factor LBD rates for BOS are estimated to be lower, with global rates at 11 % [45] and 15 % [47], with national variation in the range of 2.9–25.2 % [45]. In line with the market shift around 2008 in the silicon industry, reported by Candelise, *et al.* [55] to affect the PV module prices, several recent studies include silicon prices as explanatory variables or exclude older historical observations [21,45,53,54,56,57]. Louwen and van Sark [57] found here that the application of multi-factor learning curves reduces the impact of learning by doing from 21 % to 15.6 % in a three-factor model.

For wind power, the studies on learning rates often differentiate on onshore and offshore installations and geographical regions. Most studies address onshore installation. In their review, Rubin, Azevedo, Jaramillo and Yeh [9] find an overall range of LBD rates spanning from –11 % to 35 %, while aggregate global rates range from 8 % to 30 %. For individual European countries, the range span from 4 % to 10 % for capacity costs and -3 % to 25 % for cost of generated electricity. Samadi [16] finds lower learning rates and a narrower span on global learning rates for capacity cost (2–8 %) when reviewing studies using more recent data. The global learning rate of Williams, *et al.* [58] is estimated to 9.8 % for cost/electricity generated. Most recent one-factor studies present learning rates for US [58,59] or China [35,60], with LBD rates in the range 9–20 % and 4.98–7.50 %, respectively. All these recent studies, but Tu, Betz, Mo, Fan and Liu [60] use cost of generated electricity as explained variable. As discussed by Samadi [16], learning rates for cost of generated electricity tends to be higher than for cost of capacity as the benefits from design developments improving the capacity factor are not captured by cost of capacity rates. Furthermore, Junginger, Hittinger, Williams and Wiser [32] report even higher learning rates if accounting for quality of wind sites (6.6 % to 10.1 % to 11.4 %). Zhou and Gu [56] present a two-factor model for the United States finding a LBD rate of 17.53 % and LBR of 37.13 %. A multi-factor study for European countries [61] finds lower rates, in the range 2.34–2.62 % for LBD and 2.13–5.72 % for LBR. Both the reviews of Rubin, Azevedo, Jaramillo and Yeh [9] and

Samadi [16] found the literature on offshore wind learning rates to be scarce, and no additions to this literature has been found by the authors of this review. Rubin, Azevedo, Jaramillo and Yeh [9] report turbine learning rate estimates in the range 5–19 %, while Samadi [16] observes estimates of 0–3 %. Samadi [16] suggest the low learning rates for offshore wind turbines are partly caused by high commodity prices giving increasing turbine costs in the period of offshore installations in the early part of this century. Using data from other types of offshore installations, Junginger, *et al.* [62] estimates learning rates of interconnection cables at 38 % and HVDC converter stations at 29 % while the learning rate for erection costs is estimated to 23 % based on two installation projects [9]. Junginger, *et al.* [63] highlight differences in the learning rate between different countries and propose to use the weighted average cost of capital and water depth to obtain improved estimates of learning rates.

For electricity production from fossil fuels, *i.e.*, power plants using coal or natural gas as fuels, it is unfortunately difficult to find up-to-date empirical learning rate data. Rubin, Azevedo, Jaramillo and Yeh [9] and Samadi [16] provide references based on historical data only up until 1998 for natural gas plants and 2006 for coal plants, and most new publications on the topic appear to be based on the same sources. Other authors have also noted the lack of new learning curve parameters for these technologies that is based on actual cost and installation data [64]. For natural gas, the most relevant technology to use for extrapolation are the combined-cycle gas turbines (CCGT), since conventional open-cycle gas turbines are considered obsolete and assumed to have negligible future learning left [65]. According to Rubin, Azevedo, Jaramillo and Yeh [9], typical one-factor learning rates for this technology without CCS range from –11 % to 34 % while Samadi [16] lists –13 % to 25 %, depending on, *e.g.*, what specific measures are used as proxy for the experience and cost. As for coal plants, integrated gasification combined cycle plants (IGCC) are expected to become more important in the future, as they operate at higher efficiencies and produce lower levels of harmful emissions [65]. Furthermore, IGCC plants allow pre-combustion capture resulting in reduced costs for carbon capture. However, the empirical learning curve data found in the literature is for conventional pulverized coal (PC) power plants, as the number of IGCC power plants in operation is still very limited (Rubin, 2015). The projected learning rates for this emerging IGCC technology based on bottom-up analyses ranges from 2.5 % to 16 %, while the historical PC learning rates range from 5.6 % to 12 %, in both cases without CCS (Rubin, 2015).

Regarding energy storage, Schmidt, *et al.* [66] performed an extensive study for different electricity storage options from electronics to utility scale. Learning rates varies between –1±8 % for pumped hydro and 13±5 % for lead acid residential batteries. Most of the investigated storage options are still emerging and maturing allowing for further cost reductions while both lead acid residential batteries and pumped hydro are in a mature phase showing limiting learning potential.

**Appendix B: Learning by doing in carbon capture and storage**

Carbon capture and storage, and especially carbon capture from fossil power plants, has received considerable attention in recent years. However, there are currently only a limited number of plants installed worldwide. It is therefore not possible to use historical data to estimate learning rates, making it necessary to use surrogate technologies instead. Lohwasser and Madlener [67] used a two-factor learning curve based on flue gas desulfurization. The reason for using FDG as surrogate technology is the similarity in the process to the typical amine washing, which is a reference technology for carbon capture. Learning-by-research is accounted for through the number of patent filings as proxy whiles the installed capacity is used in learning-by-doing. The reported learning rates are 7.1 % for learning-by-doing and 6.6 % for learning-by-research. The simulation shows that the learning rates have an impact on the diffusion of carbon capture in the power market. They do however not assume learning for both transport and storage as these technologies are considered mature due to the experience from the oil and gas industry.

On the other hand, Upstill and Hall [68] investigated learning effects for $CO_2$ storage. The basis for the cost is the relative cost, that is the cost per ton $CO_2$ stored. They differentiate between four different storage sites with different associated costs and cost distributions between a $CO_2$ component and an oil and gas component to account for the difference in maturity. Hence, they utilize a composite learning curve. The learning rate for the oil and gas component is assumed to be 3 % while the learning rate for the $CO_2$ component is assumed to be 10 % based on the typical learning rates for emerging technologies. The main focus of this paper is the aggregation of the different technologies into a single learning rate accounting for varying distribution of the site selection. The overall learning rates thus vary. Reported values are in the range of 3.6-4.6 %, depending on the distribution. This approach may improve the quality of learning rates in energy system models that can be due to problem size only incorporate single storage sites.

Guo and Huang [69] utilized learning rates for the development of a CCS retrofit deployment roadmap to coal power plants in China using a mixed-integer nonlinear problem. They utilize different learning rates for variable capture and investment cost based on literature data. An important aspect of their implementation is that simultaneously build capture plants have the same costs. That implies that there is no spillover between projects at the same time and learning is only achieved by previous implementation. The variable capture cost has a learning rate of 20 %, while the investment cost has a learning rate of 5 %. The cost distribution for carbon capture hence changes due to the different learning rates with a higher share for the fixed operating costs and a reduced share for variable operating costs.

**Appendix C: Collected learning data**

In this appendix, we list the studies we encountered during our literature review for the learning rates of energy technologies. Note that we do not repeat references that are already cited in the previous literature reviews on the topic by Rubin [9] and Samadi [16]. The main results of our literature review, such as the box plot shown in Figure 2, are based on the combined data from all three sources (Rubin, Samadi, and this Appendix), of course excluding duplicate sources that were cited by both Rubin's and Samadi's papers.

There are a number of recent studies on learning effects for photovoltaics and wind power, including papers using data from up until two years ago. Tab. C1 and C2 shows what learning-by-doing (LBD) learning rates these papers have found. For the studies that used a two-factor learning curve, the learning-by-research (LBR) rate is also shown. We include the final year of data that was used to estimate these learning rates as a measure of recency, and the region to show how relevant that data is for regional energy system models. We have not found more publications on concentrated solar power (CSP) beyond what is listed in Table A4 of Samadi [16]. The table there lists 5 external references, where the last one is based on data as recent as 2013. The most recent papers cited there, *i.e.*, those that include data from after 2000, all use installed capacity as their measure of experience and investment costs as their cost variable. They then end up with LBD rates of 11–16%.

For 2$^{nd}$-generation biofuel plants, there is limited data available for estimating learning rates empirically [70]. Some estimates based on comparisons with similar technologies however exist. Recent learning data for biofuels in general is summarized in Tab. C3.

For natural gas plants, the sources listed by Rubin [9] and Samadi [16] only use data recorded before 1998, yielding learning rate information that is over 20 years out of date. We have not found any up-to-date learning rate data for natural gas plants, as all later publications we identified appear to ultimately be based on the same data sources *via* varying numbers of intermediate citations. As natural gas power plants—and in particular combined-cycle gas turbines (CCGT)—are expected to remain relevant, updated data on their learning rates would be beneficial for energy system modeling. As for carbon capture in conventional power plants, there is next to no data available regarding learning rates. The learning rates currently used to model carbon capture in general assume analogies to existing processes like flue gas desulphurization for post combustion capture with for example estimated learning rates of 7.1% for LBD and 6.6% for LBR [67].

When it comes to hydrogen as an energy carrier, it is relevant to include both learning effects related to hydrogen *usage* and hydrogen *production*. On the usage side, our review has focused on fuel cells use for domestic and mobile applications, and alternative uses have not been investigated. These results are summarized in Tab. C4, and a more detailed analysis for SOFC can be found in Ref. [71]. On the production side, Böhm et al. [19] performed a component analysis of learning rates for different types of electrolysers, yielding very recent and up-to-date estimates for the learning rates. Notably, these learning rates are expected to decrease over time due to a changing cost distribution among components. Schmidt et al. [28] compared learning rates to an expert elicitation study revealing an overlap between expert's estimation and learning rates. As for other technologies, we have not found any references on learning rates for hydrogen production with carbon capture and storage. The hydrogen production learning rates we discovered are listed in Tab. C5.

Finally, for energy storage technologies, learning rates were investigated in depth by Schmidt et al. [66]. The results are summarized in Tab. C6.

**Table C1.** Recent studies that report learning rates for solar power and specifically photovoltaics.

| Source | Cost or price | Experience | Region | End year | LBD | LBR |
|---|---|---|---|---|---|---|
| Chen, Altermatt [50] | Manufacturing cost | Cumulative production [MW] | Global | 2017 | 24% | — |
| | | | Global | 2017 | 19% | — |
| | | | Global | 2017 | 8% | — |
| Bhandari [52] | Module price | Cumulative installed capacity [MW] | Germany | 2015 | 40% | — |
| | | | Germany | 2012 | 30% | — |
| | | | Germany | 2010 | 20% | — |
| Reichelstein, Sahoo [53] | Core production cost | Production capacity [MW] | Global | 2013 | 38% | — |
| Elshurafa, Albardi [45] | Capital cost | Cumulative system installation [MW] | Global | 2015 | 11% | — |
| | | | Norway | 2014 | 8% | — |
| | | | Norway | 2014 | 17% | — |
| | | | Norway | 2014 | 7% | — |
| | | | Europe | 2013 | 17% | — |
| | | | Europe | 2013 | 9% | — |
| | | | Europe | 2013 | 9% | — |
| D'Errico [47] | Capital cost | Cumulative installed capacity [MW] | Global | 2015 | 15% | — |
| ITRPV [21] | Average module sales price | Cumulative PV module shipments [MW] | Global | 2018 | 23% | — |
| | | | Global | 2018 | 40% | — |
| Kim, Cheon [51] | Average module price | Cumulative PV production [MW] | Global | 2015 | 9% | — |
| Ding, Zhou [54] | Production cost | R&D investments | Global | 2015 | 49% | — |
| | | | Global | 2015 | 36% | — |
| | | | Germany | 2015 | 60% | — |
| | | | Germany | 2015 | 58% | — |
| Zhou, Gu [56] | Investment cost | Cumulative capacity and RD&D spending | US | 2016 | 7% | 75% |

**Table C2.** Recent studies that report learning rates for wind power.

| Source | Cost or price | Experience | Region | End year | LBD | LBR |
|---|---|---|---|---|---|---|
| Chen, Gao [35] | LCOE | Cumulative installed capacity | China | 2017 | 5% | 7% |
| Odam, de Vries [61] | Specific investment cost | Cumulative capacity, knowledge stock, scale, feed-in tariffs, commodity index | Europe | 2000 | 2% | 4% |
| Deng, Lv [72] | Unit investment cost | Cumulative installed capacity and public RD&D spending | US | 2016 | 18% | 37% |
| Wiser, Bolinger [59] | Average all-in lifetime OPEX | Global cumulative installed capacity | US | 2018 | 9% | — |
| Tu, Betz [60] | Capacity cost | Cumulative installed capacity | China | 2015 | 8% | — |
| Williams, Hittinger [58] | LCOE | Cumulative generation [kWh] | Global | 2015 | 10% | — |

**Table C3.** Learning rates for biofuel plants.

| Source | Cost or price | Experience | Region | End year | LBD | LBR |
|---|---|---|---|---|---|---|
| Daugaard, Mutti [73] | Plant costs | Cumulative production | — | — | 20% | — |
| | | | | | 5% | — |
| | Delivery costs | Cumulative production | — | — | 14% | — |
| | | | | | 10% | — |
| | Feedstock costs | Cumulative production | — | — | 14% | — |
| | | | | | 10% | — |
| de Wit, Junginger [70] | Costs | Cumulative capacity | Europe | — | 20% | — |
| | | | | | 20% | — |
| | | | | | 10% | — |
| | | | | | 2% | — |
| | | | | | 1% | — |

**Table C4.** Learning rates for hydrogen fuel cells.

| Source | Cost or price | Experience | Region | Tech | LBD | LBR |
|---|---|---|---|---|---|---|
| Staffell, Scamman [74] | Price per kW | Cumulative production [kW] | Japan | PEMFC | 16% | — |
| | | | Korea | PEMFC | 21% | — |
| | | | US | MCFC | 5% | — |
| | | | US | SOEFC | –2% | — |
| Wei, Smith [75] | Price per kW | Cumulative production [kW] | — | MCFC CHP | 4.2% | — |
| | | | — | PAFC CHP | 8.5% | — |
| | | | — | SOFC power | –1.0% | — |
| Staffell and Green [76] | Price per system | Cumulative production [kW] | EneFarm | PMFC | 15.0% | — |
| | | | Korean system | PMFC | 18.1% | — |
| | | | Anonymous | PMFC | 15.4% | — |

**Table C5.** Learning rates for hydrogen production.

| Source | Cost or price | Experience | Type | Tech | LBD | LBR |
|---|---|---|---|---|---|---|
| Böhm, Goers [19] | Manufacturing cost | Cumulative production [MW] | Electrolysers | AEC | 19.5% | — |
| | | | | PEMEC | 17.5% | — |
| | | | | SOEC | 20.5% | — |
| Schmidt, Gambhir [28] | Manufacturing cost | Cumulative production [MW] | Electrolysers | AEC | 18% | — |
| | | | | PEMEC | 18% | — |
| | | | | SOEC | 26% | — |
| Schoots, Ferioli [27] | Manufacturing cost | Cumulative production [MW] | Electrolyser | AEC | 18% | — |
| | | | SMR | SMR | 11% | — |

**Table C6.** Learning rates for energy storage technologies.

| Source | Cost or price | Experience | Type | Tech | LBD | LBR |
|---|---|---|---|---|---|---|
| Schmidt, Hawkes [66] | Manufacturing cost | Cumulative production [MW] | Pumped hydro | Utility | –1% | — |
| | | | Lead-acid | Multiple | 4% | — |
| | | | | Residential | 13% | — |
| | | | Lithium-ion | Electronics | 30% | — |
| | | | | EV | 16% | — |
| | | | | Residential | 12% | — |
| | | | | Utility | 12% | — |
| | | | NiMH | HEV | 11% | — |
| | | | V Redox flow | Utility | 11% | — |

**References**

1. Leyen, U.v.d. A Union that strives for more: My agenda for Europe. **2019**, doi:10.2775/018127.
2. Pilzecker, A.; Fernandez, R.; Mandl, N.; Rigler, E. *Annual European Union greenhouse gas inventory 1990–2018 and inventory report 2020*; European Commission, DG Climate Action, European Environment Agency: Brussles, 27.05.2020 2020; p. 997.
3. Lolou, R.; Goldstein, G.; Kanuda, A.; Lettila, A.; Remme, U. *Documentation of the TIMES Model - Part 1*; IEA Energy Technology Systems Analysis Programme: July 2016 2016; p. 151.
4. E3MLab/ICCS. *PRIMES MODEL - Detailed model description*; National Technical University of Athens: 2013-2014.
5. European Commission. *A Clean Planet for all - A European long-term strategic vision for a prosperous, modern, competitive and climate neutral economy*; European Commission: 2018.
6. Samadi, S. A Review of factors influencing the cost development of electricity generation technologies. *Energies* **2016**, *9*, doi:10.3390/en9110970.
7. Solow, R.M. A Contribution to the Theory of Economic Growth. *The Quarterly Journal of Economics* **1956**, *70*, 65-94, doi:10.2307/1884513.
8. Romer, P.M. Increasing Returns and Long-Run Growth. *Journal of Political Economy* **1986**, *94*, 1002-1037, doi:10.1086/261420.
9. Rubin, E.S.; Azevedo, I.M.L.; Jaramillo, P.; Yeh, S. A review of learning rates for electricity supply technologies. *Energy Policy* **2015**, *86*, 198-218, doi:10.1016/j.enpol.2015.06.011.
10. Rubin, E.S. Improving cost estimates for advanced low-carbon power plants. *International Journal of Greenhouse Gas Control* **2019**, *88*, 1-9, doi:10.1016/j.ijggc.2019.05.019.
11. Roussanaly, S.; Rubin, E.S.; Spek, M.v.d.; Booras, M.; Berghout, G.; Fout, N.; Garcia, T.; Gardarsdottir, M.; Kuncheekanna, S.; Matuszewski, V.N.; et al. Towards improved guidelines for cost evaluation of carbon capture and storage. *Zenodo* **2021**, doi:10.5281/zenodo.4646284.
12. Wright, T.P. Factors Affecting the Cost of Airplanes. *Journal of the Aeronautical Sciences* **1936**, *3*, 122-128, doi:10.2514/8.155.
13. Yeh, S.; Rubin, E.S. A review of uncertainties in technology experience curves. *Energy Economics* **2012**, *34*, 762-771, doi:10.1016/j.eneco.2011.11.006.
14. Heuberger, C.F.; Rubin, E.S.; Staffell, I.; Shah, N.; Dowell, N.M. Power Generation Expansion Considering Endogenous Technology Cost Learning. *Computer Aided Chemical Engineering* **2017**, *40*, 2401-2406, doi:10.1016/B978-0-444-63965-3.50402-5.
15. Rubin, E.S.; Yeh, S.; Antes, M.; Berkenpas, M.; Davison, J. Use of experience curves to estimate the future cost of power plants with CO2 capture. *International Journal of Greenhouse Gas Control* **2007**, *1*, 188-197, doi:10.1016/s1750-5836(07)00016-3.
16. Samadi, S. The experience curve theory and its application in the field of electricity generation technologies – A literature review. *Renewable and Sustainable Energy Reviews* **2018**, *82*, 2346-2364, doi:10.1016/j.rser.2017.08.077.
17. Nicodemus, J.H. Technological learning and the future of solar H2: A component learning comparison of solar thermochemical cycles and electrolysis with solar PV. *Energy Policy* **2018**, *120*, 100-109, doi:10.1016/j.enpol.2018.04.072.
18. Anandarajah, G.; McDowall, W.; Ekins, P. Decarbonising road transport with hydrogen and electricity: Long term global technology learning scenarios. *International Journal of Hydrogen Energy* **2013**, *38*, 3419-3432, doi:10.1016/j.ijhydene.2012.12.110.
19. Böhm, H.; Goers, S.; Zauner, A. Estimating future costs of power-to-gas – a component-based approach for technological learning. *International Journal of Hydrogen Energy* **2019**, *44*, 30789-30805, doi:10.1016/j.ijhydene.2019.09.230.
20. Thomassen, G.; Passel, S.v.; Dewulf, J. A review on learning effects in prospective technology assessment
. *Renewable and Sustainable Energy Reviews* **2020**, doi:10.1016/j.rser.2020.109937
21. ITRPV, I.T.R.f.P. *Results 2018 including maturity report 2019*; October 2019 2019.
22. Görig, M.; Breyer, C. Energy learning curves of PV systems. *Environmental Progress & Sustainable Energy* **2016**, *35*, 914-923, doi:10.1002/ep.12340.
23. European Commission. A hydrogen strategy for a climate-neutral Europe. **2020**.

24. Panos, E.; Kober, T. *Report on energy model analysis of the role of H2-CCS systems in Swiss energy supply and mobility with quantification of economic and environmental trade-offs, including market assessment and business case drafts*; 2020.
25. IEA. *The Future of Hydrogen*; Japan, June 2019 2019.
26. Rubin, E.S.; Yeh, S.; Antes, M.; Berkenpas, M. *Estimating the Future Trends in the Cost of CO2 Capture Technologies*; 2006/6; IEA Greenhouse Gas R&D Programme (IEAGHG): Cheltenham, UK, Februrary 2006 2006.
27. Schoots, K.; Ferioli, F.; Kramer, G.J.; van der Zwaan, B.C.C. Learning curves for hydrogen production technology: An assessment of observed cost reductions. *International Journal of Hydrogen Energy* **2008**, *33*, 2630-2645, doi:10.1016/j.ijhydene.2008.03.011.
28. Schmidt, O.; Gambhir, A.; Staffell, I.; Hawkes, A.; Nelson, J.; Few, S. Future cost and performance of water electrolysis: An expert elicitation study. *International Journal of Hydrogen Energy* **2017**, *42*, 30470-30492, doi:10.1016/j.ijhydene.2017.10.045.
29. Krishnan, S.; Fairlie, M.; Andres, P.; de Groot, T.; Jan Kramer, G. Chapter 10 - Power to gas (H2): alkaline electrolysis. In *Technological Learning in the Transition to a Low-Carbon Energy System*, Junginger, M., Louwen, A., Eds.; Academic Press: 2020; pp. 165-187.
30. Haltiwanger, J.F.; Davidson, J.H.; Wilson, E.J. Renewable hydrogen from the Zn/ZnO solar thermochemical cycle: A cost and policy analysis. In Proceedings of the ASME 2010 4th International Conference on Energy Sustainability, ES 2010, May 17, 2010 - May 22, 2010, Phoenix, AZ, United states, 2010; pp. 115-124.
31. Dutton, J.M.; Thomas, A. Treating Progress Functions as a Managerial Opportunity. *The Academy of Management Review* **1984**, *9*, 235-247.
32. Junginger, M.; Hittinger, E.; Williams, E.; Wiser, R. Chapter 6 - Onshore wind energy. In *Technological Learning in the Transition to a Low-Carbon Energy System*, Junginger, M., Louwen, A., Eds.; Academic Press: 2020; pp. 87-102.
33. Gómez, T.L.B. Technological Learning in Energy Optimisation Models and Deployment of Emerging Technologies. Eidgenössische Technische Hochschule Zürich, 2001.
34. Daggash, H.A.; Mac Dowell, N. The implications of delivering the UK's Paris Agreement commitments on the power sector. *International Journal of Greenhouse Gas Control* **2019**, *85*, 174-181, doi:10.1016/j.ijggc.2019.04.007.
35. Chen, H.; Gao, X.-Y.; Liu, J.-Y.; Zhang, Q.; Yu, S.; Kang, J.-N.; Yan, R.; Wei, Y.-M. The grid parity analysis of onshore wind power in China: A system cost perspective. *Renewable Energy* **2020**, *148*, 22-30, doi:10.1016/j.renene.2019.11.161.
36. Handayani, K.; Krozer, Y.; Filatova, T. From fossil fuels to renewables: An analysis of long-term scenarios considering technological learning. *Energy Policy* **2019**, *127*, 134-146, doi:10.1016/j.enpol.2018.11.045.
37. Cerniauskas, S.; Grube, T.; Praktiknjo, A.; Stolten, D.; Robinius, M. Future hydrogen markets for transportation and industry: The impact of CO2 taxes. *Energies* **2019**, *12*, doi:10.3390/en12244707.
38. U.S. Energy Information Administration. *The National Energy Modeling System: An Overview 2018*; U.S. Department of Energy: 2019.
39. Gumerman, E.; Marnay, C. *Learning and Cost Reductions for Generating Technologies in the National Energy Modeling System (NEMS)*; LBNL- 52559; Berkeley Lab: 2004.
40. Luderer, G.; Leimbach, M.; Bauer, N.; Kriegler, E.; Baumstark, L.; Bertram, C.; Giannousakis, A.; Hilaire, J.; Klein, D.; Levesque, A.; et al. *Description of the REMIND model (Version 1.6)*; Potsdam Institure for Climate Impact Research: Potsdan, November 2015 2015; p. 44.
41. Evans, S.; Hausfather, Z. Q&A: How 'integrated assessment models' are used to study climate change. Available online: https://www.carbonbrief.org/qa-how-integrated-assessment-models-are-used-to-study-climate-change (accessed on 2020-08-31).
42. REFLEX EU. Available online: http://reflex-project.eu/ (accessed on 2021-05-26).
43. Louwen, A.; Schreiber, S.; Junginger, M. Chapter 3 - Implementation of experience curves in energy-system models. In *Technological Learning in the Transition to a Low-Carbon Energy System*, Junginger, M., Louwen, A., Eds.; Academic Press: 2020; pp. 33-47.
44. Narbel, P.A.; Hansen, J.P. Estimating the cost of future global energy supply. *Renewable and Sustainable Energy Reviews* **2014**, *34*, 91-97, doi:10.1016/j.rser.2014.03.011.

45. Elshurafa, A.M.; Albardi, S.R.; Bigerna, S.; Bollino, C.A. Estimating the learning curve of solar PV balance–of–system for over 20 countries: Implications and policy recommendations. *Journal of Cleaner Production* **2018**, *196*, 122-134, doi:10.1016/j.jclepro.2018.06.016.
46. Viebahn, P.; Lechon, Y.; Trieb, F. The potential role of concentrated solar power (CSP) in Africa and Europe—A dynamic assessment of technology development, cost development and life cycle inventories until 2050. *Energy Policy* **2011**, *39*, 4420-4430, doi:10.1016/j.enpol.2010.09.026.
47. D'Errico, M.C. Bayesian Estimation of the Photovoltaic Balance-of-System Learning Curve. *Atlantic Economic Journal* **2019**, *47*, 111-112, doi:10.1007/s11293-019-09608-7.
48. Mauleón, I.; Hamoudi, H. Photovoltaic and wind cost decrease estimation: Implications for investment analysis. *Energy* **2017**, *137*, 1054-1065, doi:10.1016/j.energy.2017.03.109.
49. Duke, R.; Williams, R.; Payne, A. Accelerating residential PV expansion: demand analysis for competitive electricity markets. *Energy Policy* **2005**, *33*, 1912-1929, doi:10.1016/j.enpol.2004.03.005.
50. Chen, Y.; Altermatt, P.P.; Chen, D.; Zhang, X.; Xu, G.; Yang, Y.; Wang, Y.; Feng, Z.; Shen, H.; Verlinden, P.J. From Laboratory to Production: Learning Models of Efficiency and Manufacturing Cost of Industrial Crystalline Silicon and Thin-Film Photovoltaic Technologies. *IEEE Journal of Photovoltaics* **2018**, *8*, 1531-1538, doi:10.1109/jphotov.2018.2871858.
51. Kim, H.; Cheon, H.; Ahn, Y.-H.; Choi, D.G. Uncertainty quantification and scenario generation of future solar photovoltaic price for use in energy system models. *Energy* **2019**, *168*, 370-379, doi:10.1016/j.energy.2018.11.075.
52. Bhandari, R. Riding through the Experience Curve for Solar Photovoltaics Systems in Germany. In Proceedings of the 2018 7th International Energy and Sustainability Conference (IESC), 17-18 May 2018, 2018; pp. 1-7.
53. Reichelstein, S.; Sahoo, A. Relating Product Prices to Long-Run Marginal Cost: Evidence from Solar Photovoltaic Modules. *Contemporary Accounting Research* **2018**, *35*, 1464-1498, doi:10.1111/1911-3846.12319.
54. Ding, H.; Zhou, D.Q.; Liu, G.Q.; Zhou, P. Cost reduction or electricity penetration: Government R&D-induced PV development and future policy schemes. *Renewable and Sustainable Energy Reviews* **2020**, *124*, doi:10.1016/j.rser.2020.109752.
55. Candelise, C.; Winskel, M.; Gross, R.J.K. The dynamics of solar PV costs and prices as a challenge for technology forecasting. *Renewable and Sustainable Energy Reviews* **2013**, *26*, 96-107, doi:10.1016/j.rser.2013.05.012.
56. Zhou, Y.; Gu, A. Learning Curve Analysis of Wind Power and Photovoltaics Technology in US: Cost Reduction and the Importance of Research, Development and Demonstration. *Sustainability* **2019**, *11*, doi:10.3390/su11082310.
57. Louwen, A.; van Sark, W. Chapter 5 - Photovoltaic solar energy. In *Technological Learning in the Transition to a Low-Carbon Energy System*, Junginger, M., Louwen, A., Eds.; Academic Press: 2020; pp. 65-86.
58. Williams, E.; Hittinger, E.; Carvalho, R.; Williams, R. Wind power costs expected to decrease due to technological progress. *Energy Policy* **2017**, *106*, 427-435, doi:10.1016/j.enpol.2017.03.032.
59. Wiser, R.; Bolinger, M.; Lantz, E. Assessing wind power operating costs in the United States: Results from a survey of wind industry experts. *Renewable Energy Focus* **2019**, *30*, 46-57, doi:10.1016/j.ref.2019.05.003.
60. Tu, Q.; Betz, R.; Mo, J.; Fan, Y.; Liu, Y. Achieving grid parity of wind power in China – Present levelized cost of electricity and future evolution. *Applied Energy* **2019**, *250*, 1053-1064, doi:10.1016/j.apenergy.2019.05.039.
61. Odam, N.; de Vries, F.P. Innovation modelling and multi-factor learning in wind energy technology. *Energy Economics* **2020**, *85*, doi:10.1016/j.eneco.2019.104594.
62. Junginger, M.; Faaij, A.; Turkenburg, W.C. Cost Reduction Prospects for Offshore Wind Farms. *Wind Engineering* **2004**, *28*, 97-118, doi:10.1260/0309524041210847.
63. Junginger, M.; Louwen, A.; Gomez Tuya, N.; de Jager, D.; van Zuijlen, E.; Taylor, M. Chapter 7 - Offshore wind energy. In *Technological Learning in the Transition to a Low-Carbon Energy System*, Junginger, M., Louwen, A., Eds.; Academic Press: 2020; pp. 103-117.

64. Bauer, C.; Hirschberg, S.; Bäuerle, Y.; Biollaz, S.; Calbry-Muzyka, A.; Cox, B.; Heck, T.; Lehnert, M.; Meier, A.; Prasser, H.-M.; et al. *Potential, costs and environmental assessment of electricity generation technologies*; PSI, WSL, ETHZ, EPFL: Villigen, November 2017 2017; p. 783.
65. Lacal Arantegui, R.; Jaeger-Waldau, A.; Vellei, M.; Sigfusson, B.; Magagna, D.; Jakubcionis, M.; Perez Fortes, M.D.M.; Lazarou, S.; Giuntoli, J.; Weidner Ronnefeld, E.; et al. *ETRI 2014 - Energy Technology Reference Indicator projections for 2010-2050*; Joint Research Centre: 2014-12-09 2014.
66. Schmidt, O.; Hawkes, A.; Gambhir, A.; Staffell, I. The future cost of electrical energy storage based on experience rates. *Nature Energy* **2017**, *2*, doi:10.1038/nenergy.2017.110.
67. Lohwasser, R.; Madlener, R. Relating R&D and investment policies to CCS market diffusion through two-factor learning. *Energy Policy* **2013**, *52*, 439-452, doi:10.1016/j.enpol.2012.09.061.
68. Upstill, G.; Hall, P. Estimating the learning rate of a technology with multiple variants: The case of carbon storage. *Energy Policy* **2018**, *121*, 498-505, doi:10.1016/j.enpol.2018.05.017.
69. Guo, J.-X.; Huang, C. Feasible roadmap for CCS retrofit of coal-based power plants to reduce Chinese carbon emissions by 2050. *Applied Energy* **2020**, *259*, doi:10.1016/j.apenergy.2019.114112.
70. de Wit, M.; Junginger, M.; Lensink, S.; Londo, M.; Faaij, A. Competition between biofuels: Modeling technological learning and cost reductions over time. *Biomass and Bioenergy* **2010**, *34*, 203-217, doi:10.1016/j.biombioe.2009.07.012.
71. Rivera-Tinoco, R.; Schoots, K.; van der Zwaan, B. Learning curves for solid oxide fuel cells. *Energy Conversion and Management* **2012**, *57*, 86-96, doi:10.1016/j.enconman.2011.11.018.
72. Deng, X.; Lv, T. Power system planning with increasing variable renewable energy: A review of optimization models. *Journal of Cleaner Production* **2020**, *246*, doi:10.1016/j.jclepro.2019.118962.
73. Daugaard, T.; Mutti, L.A.; Wright, M.M.; Brown, R.C.; Componation, P. Learning rates and their impacts on the optimal capacities and production costs of biorefineries. *Biofuels, Bioproducts and Biorefining* **2015**, *9*, 82-94, doi:10.1002/bbb.1513.
74. Staffell, I.; Scamman, D.; Velazquez Abad, A.; Balcombe, P.; Dodds, P.E.; Ekins, P.; Shah, N.; Ward, K.R. The role of hydrogen and fuel cells in the global energy system. *Energy & Environmental Science* **2019**, *12*, 463-491, doi:10.1039/c8ee01157e.
75. Wei, M.; Smith, S.J.; Sohn, M.D. Experience curve development and cost reduction disaggregation for fuel cell markets in Japan and the US. *Applied Energy* **2017**, *191*, 346-357, doi:10.1016/j.apenergy.2017.01.056.
76. Staffell, I.; Green, R. The cost of domestic fuel cell micro-CHP systems. *International Journal of Hydrogen Energy* **2013**, *38*, 1088-1102, doi:10.1016/j.ijhydene.2012.10.090.